\documentclass[11pt]{article}

\usepackage[T1]{fontenc}
\usepackage[utf8]{inputenc}
\usepackage{lmodern}
\usepackage{geometry}
\usepackage{amsmath,amssymb,mathtools,bm}
\usepackage{physics}
\usepackage{enumitem}
\usepackage{xcolor}
\usepackage{hyperref}
\usepackage{booktabs}
\usepackage[most]{tcolorbox}
\usepackage{microtype}
\usepackage{hyperref}
\usepackage{authblk}
\hypersetup{
    colorlinks=true,
    linkcolor=blue!60!black,
    citecolor=blue!60!black,
    urlcolor=blue!60!black
}
\setlist[itemize]{topsep=3pt,itemsep=2pt,parsep=1pt}
\setlist[enumerate]{topsep=3pt,itemsep=2pt,parsep=1pt}
\numberwithin{equation}{section}

\newcommand{\avg}[1]{\left\langle #1 \right\rangle}
\renewcommand{\dd}{\mathrm{d}}
\newcommand{\dt}{\,\dd t}

\newcommand{\Strat}{\circ}
\newcommand{\D}{\mathcal{D}}

\newcommand{\Fwd}{\mathrm{F}}
\newcommand{\Rev}{\mathrm{R}}
\newcommand{\sys}{\mathrm{sys}}
\newcommand{\med}{\mathrm{m}}
\newcommand{\tot}{\mathrm{tot}}

\title{Stochastic Thermodynamics and SDE-based Generative Models}
\author[]{Yaowen ZHANG}
\affil[]{Department of Physics, The Hong Kong University of Science and Technology\\ 
         \href{mailto:yaowen.zhang@connect.ust.hk}{yaowen.zhang@connect.ust.hk}}

\date{\today}

\begin{document}
\maketitle

\begin{abstract}
SDE-based generative models, including diffusion models and the Schrödinger bridge, have found broad applications in signal processing tasks such as speech enhancement, image restoration, and time-series generation. This note presents a modeling framework for such models within the context of stochastic thermodynamics. The main results of this note are trajectory-level definitions of work, heat, and entropy production, along with a generalized Jarzynski identity and a second-law-like inequality. The proposed framework extends the original Jarzynski setup to accommodate time-dependent bath temperature and nonconservative driving forces. This thermodynamic perspective may deepen our understanding of diffusion models and the Schrödinger bridge from a nonequilibrium statistical mechanics viewpoint.
\end{abstract}

\tableofcontents

\section{Model}

\subsection{Time-dependent bath}

We consider a one-dimensional overdamped Langevin dynamics
\begin{equation}
    \dot{x}_t = \mu_t F(x_t,\lambda_t)+\zeta_t,
    \label{eq:langevin_dot}
\end{equation}
or equivalently
\begin{equation}
    \dd x_t = \mu_t F(x_t,\lambda_t)\dt + \sqrt{2D_t}\,\dd W_t.
    \label{eq:langevin_sde}
\end{equation}
Here \(x_t\) is the system coordinate, \(\lambda_t\) is an externally controlled protocol, \(\mu_t\) is the mobility, \(D_t\) is the diffusion coefficient, and \(W_t\) is a standard Wiener process.  The noise satisfies
\begin{equation}
    \avg{\zeta_t}=0,
    \qquad
    \avg{\zeta_t\zeta_{t'}}=2D_t\delta(t-t').
    \label{eq:noise}
\end{equation}
The bath temperature is allowed to depend on time:
\begin{equation}
    T_t=T(t),
    \qquad
    \beta_t=\frac{1}{T_t}.
    \label{eq:beta}
\end{equation}
We set \(k_B=1\).  If one keeps Boltzmann's constant explicitly, then \(\beta_t=1/(k_BT_t)\).

The instantaneous Einstein relation is imposed:
\begin{equation}
    D_t=\mu_tT_t=\frac{\mu_t}{\beta_t}.
    \label{eq:einstein}
\end{equation}
This condition states that the stochastic forcing at time \(t\) corresponds locally to a bath with temperature \(T_t\).  In the present note, \(T_t\), \(D_t\), and \(\mu_t\) depend on time only and not on position.  Therefore the noise is additive in the stochastic-calculus sense, and the Langevin equation itself has no Ito-Stratonovich drift ambiguity.  However, the thermodynamic line integrals below do require a convention.

\subsection{Force decomposition}

The total force is decomposed into a conservative part and a nonconservative part:
\begin{equation}
    F(x,\lambda_t)=-\partial_x V(x,\lambda_t)+f(x,\lambda_t).
    \label{eq:force_decomp}
\end{equation}
Here \(V(x,\lambda)\) is the potential energy, while \(f(x,\lambda)\) is an externally applied nonconservative force.  The corresponding Fokker-Planck equation is
\begin{equation}
    \partial_t p(x,t)=-\partial_x j(x,t),
    \label{eq:fpe}
\end{equation}
with probability current
\begin{equation}
    j(x,t)=\mu_tF(x,\lambda_t)p(x,t)-D_t\partial_xp(x,t).
    \label{eq:current}
\end{equation}

\section{Trajectory-level energy, work, and heat}

\subsection{Stratonovich convention}

All thermodynamic line integrals involving \(\dd x_t\) are interpreted in the Stratonovich sense, denoted by \(\Strat\dd x_t\).  This convention is used because it preserves the ordinary chain rule.  In particular,
\begin{equation}
    \dd V(x_t,\lambda_t)
    =\partial_xV(x_t,\lambda_t)\Strat\dd x_t
    +\partial_\lambda V(x_t,\lambda_t)\dot{\lambda}_t\dt .
    \label{eq:chain_rule}
\end{equation}
This is the stochastic version of the usual differential of a function of \(x_t\) and \(\lambda_t\).

\subsection{Internal energy}

The trajectory-level internal energy is defined as
\begin{equation}
    E_t=V(x_t,\lambda_t).
    \label{eq:energy}
\end{equation}
Therefore the change in internal energy is
\begin{equation}
    \dd E_t
    =\partial_xV(x_t,\lambda_t)\Strat\dd x_t
    +\partial_\lambda V(x_t,\lambda_t)\dot{\lambda}_t\dt .
    \label{eq:dE}
\end{equation}

\subsection{Work increment}

The total work increment contains two contributions.  The first is the work due to changing the external parameter \(\lambda_t\) in the potential:
\begin{equation}
    \delta w_{\lambda,t}
    =\partial_\lambda V(x_t,\lambda_t)\dot{\lambda}_t\dt .
    \label{eq:lambda_work}
\end{equation}
The second is the work performed by the nonconservative force:
\begin{equation}
    \delta w_{f,t}=f(x_t,\lambda_t)\Strat\dd x_t .
    \label{eq:f_work}
\end{equation}
Hence the total work increment is
\begin{equation}
    \boxed{
    \delta w_t
    =\partial_\lambda V(x_t,\lambda_t)\dot{\lambda}_t\dt
    +f(x_t,\lambda_t)\Strat\dd x_t .
    }
    \label{eq:work_increment}
\end{equation}
The corresponding trajectory work functional is
\begin{equation}
    \boxed{
    W[x]
    =\int_0^\tau \partial_\lambda V(x_t,\lambda_t)\dot{\lambda}_t\dt
    +\int_0^\tau f(x_t,\lambda_t)\Strat\dd x_t .
    }
    \label{eq:work_functional}
\end{equation}

\subsection{Heat dissipated into the medium}

We use the sign convention that \(\delta q_t\) is heat dissipated from the system into the surrounding medium.  In an overdamped Langevin system, force times displacement gives the dissipated heat:
\begin{equation}
    \boxed{
    \delta q_t=F(x_t,\lambda_t)\Strat\dd x_t .
    }
    \label{eq:heat_increment}
\end{equation}
Using eq.\ref{eq:force_decomp}, this can be written explicitly as
\begin{equation}
    \delta q_t
    =\left[-\partial_xV(x_t,\lambda_t)+f(x_t,\lambda_t)\right]\Strat\dd x_t .
    \label{eq:heat_decomp}
\end{equation}
The total heat dissipated into the medium along a trajectory is
\begin{equation}
    \boxed{
    Q[x]=\int_0^\tau F(x_t,\lambda_t)\Strat\dd x_t .
    }
    \label{eq:heat_functional}
\end{equation}

\subsection{First law along a trajectory}

Starting from \eqref{eq:dE} and \eqref{eq:heat_decomp}, we compute
\begin{align}
    \dd E_t+\delta q_t
    &=\left[\partial_xV\Strat\dd x_t
      +\partial_\lambda V\dot{\lambda}_t\dt\right]
      +\left[-\partial_xV+f\right]\Strat\dd x_t \\
    &=\partial_xV\Strat\dd x_t
      -\partial_xV\Strat\dd x_t
      +\partial_\lambda V\dot{\lambda}_t\dt
      +f\Strat\dd x_t \\
    &=\partial_\lambda V\dot{\lambda}_t\dt
      +f\Strat\dd x_t \\
    &=\delta w_t .
\end{align}
Therefore
\begin{equation}
    \boxed{
    \delta w_t=\dd E_t+\delta q_t .
    }
    \label{eq:first_law}
\end{equation}
After integration from \(0\) to \(\tau\),
\begin{equation}
    \boxed{
    W[x]=\Delta E[x]+Q[x]
    }
    \label{eq:first_law_integrated}
\end{equation}
with
\begin{equation}
    \Delta E[x]=V(x_\tau,\lambda_\tau)-V(x_0,\lambda_0).
    \label{eq:deltaE}
\end{equation}
Equivalently,
\begin{equation}
    Q[x]=W[x]-\Delta E[x].
    \label{eq:q_w_e}
\end{equation}

\section{Entropy production}

\subsection{System entropy}

Let \(p(x,t)\) be the solution of the Fokker-Planck equation \eqref{eq:fpe} generated by the specified initial distribution \(p_0(x)\).  The stochastic system entropy along a trajectory is defined by
\begin{equation}
    \boxed{
    s_{\sys}(t)=-\ln p(x_t,t).
    }
    \label{eq:system_entropy}
\end{equation}
The system entropy change along the trajectory is therefore
\begin{align}
    \Delta s_{\sys}[x]
    &=s_{\sys}(\tau)-s_{\sys}(0) \\
    &=-\ln p(x_\tau,\tau)+\ln p_0(x_0).
    \label{eq:delta_system_entropy}
\end{align}

The ensemble entropy is recovered by averaging \eqref{eq:system_entropy} over \(p(x,t)\):
\begin{equation}
    S_{\sys}(t)=\avg{s_{\sys}(t)}
    =-\int \dd x\,p(x,t)\ln p(x,t).
    \label{eq:ensemble_entropy}
\end{equation}

\subsection{Medium entropy}

For a time-dependent bath temperature, the entropy change of the medium is obtained by weighting the heat increment by the instantaneous inverse temperature:
\begin{equation}
    \boxed{
    \dd s_{\med,t}=\beta_t\delta q_t
    =\beta_tF(x_t,\lambda_t)\Strat\dd x_t .
    }
    \label{eq:medium_entropy_increment}
\end{equation}
Thus
\begin{equation}
    \boxed{
    \Delta s_{\med}[x]
    =\int_0^\tau \beta_tF(x_t,\lambda_t)\Strat\dd x_t .
    }
    \label{eq:medium_entropy}
\end{equation}
Using \eqref{eq:q_w_e}, this may also be written as
\begin{equation}
    \Delta s_{\med}[x]
    =\int_0^\tau \beta_t\delta w_t
     -\int_0^\tau \beta_t\dd E_t .
    \label{eq:smed_work_energy}
\end{equation}
The second term will be used below in the Jarzynski derivation.

\subsection{Total entropy production}

The total entropy production is the sum of medium entropy and system entropy:
\begin{equation}
    \boxed{
    \Delta s_{\tot}[x]
    =\Delta s_{\med}[x]+\Delta s_{\sys}[x].
    }
    \label{eq:total_entropy}
\end{equation}
Substituting \eqref{eq:medium_entropy} and \eqref{eq:delta_system_entropy},
\begin{equation}
    \boxed{
    \Delta s_{\tot}[x]
    =\int_0^\tau \beta_tF(x_t,\lambda_t)\Strat\dd x_t
    -\ln p(x_\tau,\tau)+\ln p_0(x_0).
    }
    \label{eq:total_entropy_explicit}
\end{equation}

\subsection{Average entropy production rate}

The ensemble entropy production rate remains nonnegative.  To show this explicitly, first write the current \eqref{eq:current} as
\begin{equation}
    j=\mu_tFp-D_t\partial_xp.
    \label{eq:current_again}
\end{equation}
Solving for \(F\), and using \(D_t=\mu_tT_t\), gives
\begin{align}
    \mu_tFp &= j+D_t\partial_xp, \\
    F &= \frac{j}{\mu_tp}+\frac{D_t}{\mu_t}\partial_x\ln p, \\
    \beta_tF &= \frac{\beta_t}{\mu_t}\frac{j}{p}+\beta_t\frac{D_t}{\mu_t}\partial_x\ln p, \\
    \beta_tF &= \frac{j}{D_tp}+\partial_x\ln p.
    \label{eq:betaF_identity}
\end{align}
The average medium entropy production rate is
\begin{equation}
    \dot S_{\med}(t)
    =\int \dd x\,p(x,t)\,\beta_tF(x,\lambda_t)\avg{\dot x_t\mid x_t=x}.
    \label{eq:average_medium_start}
\end{equation}
For overdamped diffusion, the conditional mean velocity is
\begin{equation}
    \avg{\dot x_t\mid x_t=x}=\frac{j(x,t)}{p(x,t)}.
    \label{eq:conditional_velocity}
\end{equation}
Therefore
\begin{align}
    \dot S_{\med}(t)
    &=\int \dd x\,\beta_tF(x,\lambda_t)j(x,t) \\
    &=\int \dd x\,j(x,t)
      \left[\frac{j(x,t)}{D_tp(x,t)}+\partial_x\ln p(x,t)\right] \\
    &=\int \dd x\,\frac{j(x,t)^2}{D_tp(x,t)}
      +\int \dd x\,j(x,t)\partial_x\ln p(x,t).
    \label{eq:smed_rate_expand}
\end{align}
The rate of change of the ensemble system entropy is
\begin{align}
    \dot S_{\sys}(t)
    &=-\int \dd x\,\partial_tp(x,t)\ln p(x,t)
      -\int \dd x\,\partial_tp(x,t) \\
    &=-\int \dd x\,\partial_tp(x,t)\ln p(x,t).
    \label{eq:ssys_rate_1}
\end{align}
The second integral vanishes by probability conservation.  Using \(\partial_tp=-\partial_xj\),
\begin{align}
    \dot S_{\sys}(t)
    &=\int \dd x\,\partial_xj(x,t)\ln p(x,t) \\
    &=-\int \dd x\,j(x,t)\partial_x\ln p(x,t),
    \label{eq:ssys_rate_2}
\end{align}
where boundary terms are assumed to vanish.  Adding \eqref{eq:smed_rate_expand} and \eqref{eq:ssys_rate_2} gives
\begin{equation}
    \boxed{
    \dot S_{\tot}(t)
    =\dot S_{\med}(t)+\dot S_{\sys}(t)
    =\int \dd x\,\frac{j(x,t)^2}{D_tp(x,t)}\ge 0 .
    }
    \label{eq:positive_entropy_rate}
\end{equation}

\section{Path probability ratio and integral fluctuation theorem}

\subsection{Forward and reverse protocols}

The forward protocol is
\begin{equation}
    \Lambda=\{\lambda_t,\beta_t,\mu_t\}_{0\le t\le\tau}.
    \label{eq:forward_protocol}
\end{equation}
The reversed protocol is
\begin{equation}
    \Lambda^\dagger=\{\lambda^\dagger_t,\beta^\dagger_t,\mu^\dagger_t\}_{0\le t\le\tau}
    =\{\lambda_{\tau-t},\beta_{\tau-t},\mu_{\tau-t}\}_{0\le t\le\tau}.
    \label{eq:reverse_protocol}
\end{equation}
Given a forward path \(x_t\), the time-reversed path is
\begin{equation}
    x_t^\dagger=x_{\tau-t}.
    \label{eq:reverse_path}
\end{equation}
The reversed experiment starts from a distribution \(p_1(x)\), which can be chosen freely as long as it is normalized.

\subsection{Conditional path probability ratio}

For the overdamped dynamics with time-dependent but position-independent diffusion, the Onsager-Machlup path weight in the Stratonovich discretization has the form
\begin{equation}
    \mathcal{P}_{\Fwd}[x|x_0]
    \propto
    \exp\left\{-\int_0^\tau\dt
    \left[
    \frac{\left(\dot x_t-\mu_tF(x_t,\lambda_t)\right)^2}{4D_t}
    +\frac{\mu_t}{2}\partial_xF(x_t,\lambda_t)
    \right]\right\}.
    \label{eq:path_weight_forward}
\end{equation}
The reversed path has velocity \(\dot x_t^\dagger=-\dot x_{\tau-t}\).  When the reversed protocol is used, the even part of the Onsager-Machlup action is the same under the change of variable \(t\mapsto\tau-t\).  The term responsible for the ratio is obtained by expanding the square.  Locally in time,
\begin{align}
    &\frac{(\dot x_t+\mu_tF)^2-(\dot x_t-\mu_tF)^2}{4D_t} \\
    &=\frac{\dot x_t^2+2\mu_tF\dot x_t+\mu_t^2F^2
      -\dot x_t^2+2\mu_tF\dot x_t-\mu_t^2F^2}{4D_t} \\
    &=\frac{\mu_tF\dot x_t}{D_t}.
    \label{eq:square_difference}
\end{align}
Using \(D_t=\mu_t/\beta_t\),
\begin{equation}
    \frac{\mu_tF\dot x_t}{D_t}=\beta_tF\dot x_t.
    \label{eq:local_db}
\end{equation}
Therefore the conditional path probability ratio is
\begin{equation}
    \boxed{
    \ln\frac{\mathcal{P}_{\Fwd}[x|x_0]}
            {\mathcal{P}_{\Rev}[x^\dagger|x_\tau]}
    =\int_0^\tau \beta_tF(x_t,\lambda_t)\Strat\dd x_t
    =\Delta s_{\med}[x].
    }
    \label{eq:conditional_path_ratio}
\end{equation}
This is the local detailed-balance statement for the time-dependent bath.

\subsection{Total path probability ratio}

The full forward path probability includes the forward initial distribution:
\begin{equation}
    \mathbb{P}_{\Fwd}[x]
    =p_0(x_0)\mathcal{P}_{\Fwd}[x|x_0].
    \label{eq:full_forward_path}
\end{equation}
The full reversed path probability includes the reversed initial distribution \(p_1(x_\tau)\):
\begin{equation}
    \mathbb{P}_{\Rev}[x^\dagger]
    =p_1(x_\tau)\mathcal{P}_{\Rev}[x^\dagger|x_\tau].
    \label{eq:full_reverse_path}
\end{equation}
Combining \eqref{eq:conditional_path_ratio}, \eqref{eq:full_forward_path}, and \eqref{eq:full_reverse_path},
\begin{align}
    \ln\frac{\mathbb{P}_{\Fwd}[x]}{\mathbb{P}_{\Rev}[x^\dagger]}
    &=\ln\frac{p_0(x_0)}{p_1(x_\tau)}
      +\ln\frac{\mathcal{P}_{\Fwd}[x|x_0]}
               {\mathcal{P}_{\Rev}[x^\dagger|x_\tau]} \\
    &=\ln\frac{p_0(x_0)}{p_1(x_\tau)}+\Delta s_{\med}[x].
    \label{eq:full_path_ratio_general}
\end{align}
If \(p_1(x)=p(x,\tau)\), then
\begin{equation}
    \ln\frac{p_0(x_0)}{p_1(x_\tau)}
    =\ln p_0(x_0)-\ln p(x_\tau,\tau)
    =\Delta s_{\sys}[x].
    \label{eq:system_entropy_from_ratio}
\end{equation}
Thus
\begin{equation}
    \boxed{
    \ln\frac{\mathbb{P}_{\Fwd}[x]}{\mathbb{P}_{\Rev}[x^\dagger]}
    =\Delta s_{\tot}[x]
    }
    \label{eq:total_path_ratio}
\end{equation}
for the special choice \(p_1=p(\cdot,\tau)\).

\subsection{Integral fluctuation theorem}

For a general normalized \(p_1\), equation \eqref{eq:full_path_ratio_general} gives
\begin{equation}
    \frac{\mathbb{P}_{\Rev}[x^\dagger]}{\mathbb{P}_{\Fwd}[x]}
    =\exp[-\Delta s_{\med}[x]]\frac{p_1(x_\tau)}{p_0(x_0)}.
    \label{eq:ratio_exp}
\end{equation}
Averaging the right-hand side over the forward process gives
\begin{align}
    \avg{\exp[-\Delta s_{\med}[x]]\frac{p_1(x_\tau)}{p_0(x_0)}}_{\Fwd}
    &=\int \D x\,\mathbb{P}_{\Fwd}[x]
      \exp[-\Delta s_{\med}[x]]\frac{p_1(x_\tau)}{p_0(x_0)} \\
    &=\int \D x\,\mathbb{P}_{\Rev}[x^\dagger] \\
    &=\int \D x^\dagger\,\mathbb{P}_{\Rev}[x^\dagger] \\
    &=1.
    \label{eq:ift_general_derivation}
\end{align}
Therefore
\begin{equation}
    \boxed{
    \avg{\exp[-\Delta s_{\med}[x]]\frac{p_1(x_\tau)}{p_0(x_0)}}_{\Fwd}=1 .
    }
    \label{eq:general_ift}
\end{equation}
Choosing \(p_1(x)=p(x,\tau)\) yields the total entropy production identity
\begin{equation}
    \boxed{
    \avg{e^{-\Delta s_{\tot}}}=1 .
    }
    \label{eq:ift_total}
\end{equation}

\section{Generalized Jarzynski identity}

\subsection{Canonical initial and reference final distributions}

We now choose canonical distributions associated with the conservative potential \(V\).  The forward initial distribution is
\begin{equation}
    p_0(x)=\frac{e^{-\beta_0V(x,\lambda_0)}}{Z_0},
    \qquad
    Z_0=Z(\beta_0,\lambda_0)
    =\int \dd x\,e^{-\beta_0V(x,\lambda_0)}.
    \label{eq:p0_canonical}
\end{equation}
The reference final distribution is
\begin{equation}
    p_1(x)=\frac{e^{-\beta_\tau V(x,\lambda_\tau)}}{Z_\tau},
    \qquad
    Z_\tau=Z(\beta_\tau,\lambda_\tau)
    =\int \dd x\,e^{-\beta_\tau V(x,\lambda_\tau)}.
    \label{eq:p1_canonical}
\end{equation}
The corresponding equilibrium free energy is
\begin{equation}
    F(\beta,\lambda)=-\frac{1}{\beta}\ln Z(\beta,\lambda),
    \label{eq:free_energy}
\end{equation}
and the dimensionless free energy is
\begin{equation}
    \beta F(\beta,\lambda)=-\ln Z(\beta,\lambda).
    \label{eq:dimensionless_free_energy}
\end{equation}

The nonconservative force \(f\) is allowed in the trajectory dynamics and in the work functional.  If \(f\) is genuinely nonconservative, the right-hand side below is the free energy difference of the reference conservative potential \(V\), because a nonconservative force cannot generally be absorbed into a scalar equilibrium potential.

\subsection{Substitution into the general fluctuation identity}

The general identity \eqref{eq:general_ift} is
\begin{equation}
    \avg{\exp[-\Delta s_{\med}[x]]\frac{p_1(x_\tau)}{p_0(x_0)}}=1.
    \label{eq:ift_for_jarzynski}
\end{equation}
Using \eqref{eq:p0_canonical} and \eqref{eq:p1_canonical},
\begin{align}
    \frac{p_1(x_\tau)}{p_0(x_0)}
    &=\frac{e^{-\beta_\tau V(x_\tau,\lambda_\tau)}/Z_\tau}
            {e^{-\beta_0V(x_0,\lambda_0)}/Z_0} \\
    &=\frac{Z_0}{Z_\tau}
      \exp\left[-\beta_\tau V(x_\tau,\lambda_\tau)
      +\beta_0V(x_0,\lambda_0)\right].
    \label{eq:p1_over_p0}
\end{align}
For compactness, define
\begin{equation}
    V_t=V(x_t,\lambda_t).
    \label{eq:Vt_def}
\end{equation}
Then \eqref{eq:p1_over_p0} becomes
\begin{equation}
    \frac{p_1(x_\tau)}{p_0(x_0)}
    =\frac{Z_0}{Z_\tau}\exp[-\beta_\tau V_\tau+\beta_0V_0].
    \label{eq:p1_over_p0_short}
\end{equation}
Substituting into \eqref{eq:ift_for_jarzynski},
\begin{equation}
    1=\frac{Z_0}{Z_\tau}
    \avg{\exp\left[-\Delta s_{\med}[x]-\beta_\tau V_\tau+\beta_0V_0\right]}.
    \label{eq:jarzynski_intermediate_1}
\end{equation}

\subsection{Expressing medium entropy through work}

From the trajectory first law \eqref{eq:first_law},
\begin{equation}
    \delta q_t=\delta w_t-\dd E_t.
    \label{eq:q_equals_w_minus_E}
\end{equation}
Since \(E_t=V_t\),
\begin{equation}
    \delta q_t=\delta w_t-\dd V_t.
    \label{eq:q_equals_w_minus_V}
\end{equation}
Multiplying by \(\beta_t\) and integrating,
\begin{align}
    \Delta s_{\med}[x]
    &=\int_0^\tau\beta_t\delta q_t \\
    &=\int_0^\tau\beta_t\delta w_t
      -\int_0^\tau\beta_t\dd V_t.
    \label{eq:smed_work_step}
\end{align}
The second integral is evaluated by integration by parts.  Since \(\beta_t\) is an externally prescribed differentiable function of time and \(V_t\) is interpreted with the Stratonovich chain rule,
\begin{equation}
    \dd(\beta_tV_t)=\dot\beta_tV_t\dt+\beta_t\dd V_t.
    \label{eq:product_rule_betaV}
\end{equation}
Therefore
\begin{equation}
    \beta_t\dd V_t=\dd(\beta_tV_t)-\dot\beta_tV_t\dt.
    \label{eq:beta_dV_identity}
\end{equation}
Integrating over time,
\begin{align}
    \int_0^\tau\beta_t\dd V_t
    &=\int_0^\tau\dd(\beta_tV_t)
      -\int_0^\tau\dot\beta_tV_t\dt \\
    &=\beta_\tau V_\tau-\beta_0V_0
      -\int_0^\tau\dot\beta_tV_t\dt .
    \label{eq:int_beta_dV}
\end{align}
Substituting \eqref{eq:int_beta_dV} into \eqref{eq:smed_work_step},
\begin{equation}
    \Delta s_{\med}[x]
    =\int_0^\tau\beta_t\delta w_t
     -\beta_\tau V_\tau+\beta_0V_0
     +\int_0^\tau\dot\beta_tV_t\dt .
    \label{eq:smed_work_final}
\end{equation}
Hence
\begin{align}
    &-\Delta s_{\med}[x]-\beta_\tau V_\tau+\beta_0V_0 \\
    &=-\left[\int_0^\tau\beta_t\delta w_t
     -\beta_\tau V_\tau+\beta_0V_0
     +\int_0^\tau\dot\beta_tV_t\dt\right]
     -\beta_\tau V_\tau+\beta_0V_0 \\
    &=-\int_0^\tau\beta_t\delta w_t
      +\beta_\tau V_\tau-\beta_0V_0
      -\int_0^\tau\dot\beta_tV_t\dt
      -\beta_\tau V_\tau+\beta_0V_0 \\
    &=-\int_0^\tau\beta_t\delta w_t
      -\int_0^\tau\dot\beta_tV_t\dt .
    \label{eq:boundary_cancel}
\end{align}
Substituting \eqref{eq:boundary_cancel} into \eqref{eq:jarzynski_intermediate_1} gives
\begin{equation}
    1=\frac{Z_0}{Z_\tau}
    \avg{\exp\left[-\int_0^\tau\beta_t\delta w_t
    -\int_0^\tau\dot\beta_tV_t\dt\right]}.
    \label{eq:jarzynski_intermediate_2}
\end{equation}

\subsection{Reduced work functional}

Using the definition of \(\delta w_t\) in \eqref{eq:work_increment},
\begin{align}
    \int_0^\tau\beta_t\delta w_t
    &=\int_0^\tau\beta_t\partial_\lambda V(x_t,\lambda_t)\dot\lambda_t\dt
      +\int_0^\tau\beta_tf(x_t,\lambda_t)\Strat\dd x_t .
    \label{eq:beta_work_expand}
\end{align}
Define the dimensionless work-like functional
\begin{equation}
    \boxed{
    Y[x]
    =\int_0^\tau\beta_t\partial_\lambda V(x_t,\lambda_t)\dot\lambda_t\dt
    +\int_0^\tau\beta_tf(x_t,\lambda_t)\Strat\dd x_t
    +\int_0^\tau\dot\beta_tV(x_t,\lambda_t)\dt .
    }
    \label{eq:reduced_work_functional}
\end{equation}
Equivalently,
\begin{equation}
    Y[x]=\int_0^\tau\beta_t\delta w_t+
    \int_0^\tau\dot\beta_tV_t\dt .
    \label{eq:Y_compact}
\end{equation}
Then \eqref{eq:jarzynski_intermediate_2} becomes
\begin{equation}
    \boxed{
    \avg{e^{-Y[x]}}=\frac{Z_\tau}{Z_0}.
    }
    \label{eq:jarzynski_Z}
\end{equation}
Since \(\beta F=-\ln Z\),
\begin{align}
    \frac{Z_\tau}{Z_0}
    &=\exp[\ln Z_\tau-\ln Z_0] \\
    &=\exp[-\beta_\tau F(\beta_\tau,\lambda_\tau)
             +\beta_0F(\beta_0,\lambda_0)].
    \label{eq:Z_to_free_energy}
\end{align}
Therefore the generalized Jarzynski identity is
\begin{equation}
    \boxed{
    \avg{e^{-Y[x]}}
    =\exp\left\{-\left[
    \beta_\tau F(\beta_\tau,\lambda_\tau)
    -\beta_0F(\beta_0,\lambda_0)
    \right]\right\}.
    }
    \label{eq:generalized_JE}
\end{equation}

\subsection{Useful limiting forms}

If the bath temperature is constant, then \(\beta_t=\beta\) and \(\dot\beta_t=0\).  The functional \eqref{eq:reduced_work_functional} becomes
\begin{equation}
    Y[x]=\beta\left[
    \int_0^\tau\partial_\lambda V(x_t,\lambda_t)\dot\lambda_t\dt
    +\int_0^\tau f(x_t,\lambda_t)\Strat\dd x_t
    \right]
    =\beta W[x].
    \label{eq:constant_temp_Y}
\end{equation}
Therefore
\begin{equation}
    \boxed{
    \avg{e^{-\beta W[x]}}=e^{-\beta\Delta F}
    }
    \label{eq:standard_JE_with_total_work}
\end{equation}
with
\begin{equation}
    \Delta F=F(\beta,\lambda_\tau)-F(\beta,\lambda_0).
    \label{eq:deltaF_constant_beta}
\end{equation}
When \(f=0\), this is the standard equilibrium free-energy Jarzynski equality.  When \(f\neq0\), the same formal identity uses the total work including the nonconservative contribution, while the free energy on the right-hand side is the reference free energy generated by the conservative potential \(V\).

If there is no nonconservative force, then \(f=0\) and
\begin{equation}
    Y[x]
    =\int_0^\tau\beta_t\partial_\lambda V(x_t,\lambda_t)\dot\lambda_t\dt
    +\int_0^\tau\dot\beta_tV(x_t,\lambda_t)\dt .
    \label{eq:Y_no_f}
\end{equation}
If, in addition, \(\lambda_t\) is fixed and only the temperature changes, then
\begin{equation}
    Y[x]=\int_0^\tau\dot\beta_tV(x_t,\lambda)\dt,
    \label{eq:Y_temperature_only}
\end{equation}
and
\begin{equation}
    \boxed{
    \avg{\exp\left[-\int_0^\tau\dot\beta_tV(x_t,\lambda)\dt\right]}
    =\frac{Z(\beta_\tau,\lambda)}{Z(\beta_0,\lambda)} .
    }
    \label{eq:temperature_only_identity}
\end{equation}

\section{Relation between work, free energy, and entropy}
\label{sec:work_free_entropy_relation}

This section makes explicit the relation among the reduced work functional, the reference free-energy difference, and entropy production.  We keep the same assumptions as in the Jarzynski identity above: the initial distribution is the canonical distribution generated by the conservative potential at \((\beta_0,\lambda_0)\), and the reference final distribution is the canonical distribution generated by the same conservative potential at \((\beta_\tau,\lambda_\tau)\).  The actual final distribution generated by the forward dynamics is denoted by
\begin{equation}
    p_\tau(x)\equiv p(x,\tau).
\end{equation}
It need not be identical to the reference final canonical distribution
\begin{equation}
    p_{\mathrm{eq}}^\tau(x)
    =\frac{e^{-\beta_\tau V(x,\lambda_\tau)}}{Z_\tau}.
    \label{eq:peq_tau_relation_section}
\end{equation}

\subsection{Dissipated reduced work}

The generalized Jarzynski identity can be written as
\begin{equation}
    \avg{e^{-Y[x]}}=e^{-\Delta(\beta F)},
    \label{eq:JE_relation_section}
\end{equation}
where
\begin{equation}
    \Delta(\beta F)
    \equiv \beta_\tau F(\beta_\tau,\lambda_\tau)
    -\beta_0F(\beta_0,\lambda_0)
    =-\ln Z_\tau+\ln Z_0.
    \label{eq:delta_betaF_relation_section}
\end{equation}
It is therefore natural to define the dissipated reduced work
\begin{equation}
    \boxed{
    \Sigma[x]\equiv Y[x]-\Delta(\beta F).
    }
    \label{eq:Sigma_def}
\end{equation}
Then the Jarzynski identity becomes the integral fluctuation relation
\begin{equation}
    \boxed{
    \avg{e^{-\Sigma[x]}}=1.
    }
    \label{eq:Sigma_IFT}
\end{equation}
By Jensen's inequality,
\begin{equation}
    \avg{\Sigma}\geq 0.
    \label{eq:Sigma_second_law}
\end{equation}
Thus \(\Sigma\) is the reduced excess work beyond the reduced free-energy difference.

\subsection{Exact trajectory relation to total entropy production}

We now derive the relation between \(\Sigma[x]\) and the Seifert total entropy production.  From the first law,
\begin{equation}
    \delta q_t=\delta w_t-\dd V_t.
\end{equation}
Multiplying by \(\beta_t\) and integrating gives
\begin{align}
    \Delta s_{\med}[x]
    &=\int_0^\tau \beta_t\delta q_t \\
    &=\int_0^\tau \beta_t\delta w_t
      -\int_0^\tau \beta_t\dd V_t .
\end{align}
Using the Stratonovich product rule
\begin{equation}
    \dd(\beta_tV_t)=\dot\beta_tV_t\dt+\beta_t\dd V_t,
\end{equation}
we have
\begin{equation}
    \int_0^\tau \beta_t\dd V_t
    =\beta_\tau V_\tau-\beta_0V_0
    -\int_0^\tau \dot\beta_tV_t\dt .
\end{equation}
Therefore
\begin{align}
    \Delta s_{\med}[x]
    &=\int_0^\tau \beta_t\delta w_t
      -\beta_\tau V_\tau+\beta_0V_0
      +\int_0^\tau \dot\beta_tV_t\dt \\
    &=Y[x]-\beta_\tau V_\tau+\beta_0V_0 .
    \label{eq:smed_Y_relation}
\end{align}
The system entropy change is
\begin{equation}
    \Delta s_{\sys}[x]
    =-\ln p_\tau(x_\tau)+\ln p_0(x_0).
\end{equation}
Because the initial distribution is canonical,
\begin{equation}
    p_0(x_0)=\frac{e^{-\beta_0V_0}}{Z_0},
    \qquad
    \ln p_0(x_0)=-\beta_0V_0-\ln Z_0.
    \label{eq:ln_p0_canonical_relation}
\end{equation}
Thus the total entropy production is
\begin{align}
    \Delta s_{\tot}[x]
    &=\Delta s_{\med}[x]+\Delta s_{\sys}[x] \\
    &=Y[x]-\beta_\tau V_\tau+\beta_0V_0
      -\ln p_\tau(x_\tau)-\beta_0V_0-\ln Z_0 \\
    &=Y[x]-\beta_\tau V_\tau-\ln p_\tau(x_\tau)-\ln Z_0 .
    \label{eq:stot_Y_intermediate}
\end{align}
From the reference final canonical distribution \eqref{eq:peq_tau_relation_section},
\begin{equation}
    \ln p_{\mathrm{eq}}^\tau(x_\tau)
    =-\beta_\tau V_\tau-\ln Z_\tau.
    \label{eq:ln_peq_tau_relation}
\end{equation}
Hence
\begin{equation}
    \ln\frac{p_\tau(x_\tau)}{p_{\mathrm{eq}}^\tau(x_\tau)}
    =\ln p_\tau(x_\tau)+\beta_\tau V_\tau+
    \ln Z_\tau .
    \label{eq:final_mismatch_log}
\end{equation}
Adding \eqref{eq:final_mismatch_log} to \eqref{eq:stot_Y_intermediate},
\begin{align}
    \Delta s_{\tot}[x]
    +\ln\frac{p_\tau(x_\tau)}{p_{\mathrm{eq}}^\tau(x_\tau)}
    &=Y[x]-\ln Z_0+
    \ln Z_\tau \\
    &=Y[x]-\left(-\ln Z_\tau+
    \ln Z_0\right) \\
    &=Y[x]-\Delta(\beta F).
\end{align}
Therefore
\begin{equation}
    \boxed{
    \Sigma[x]
    =Y[x]-\Delta(\beta F)
    =\Delta s_{\tot}[x]
    +\ln\frac{p_\tau(x_\tau)}{p_{\mathrm{eq}}^\tau(x_\tau)} .
    }
    \label{eq:Sigma_entropy_exact}
\end{equation}
This is the desired trajectory-level equality among the work functional, the free-energy difference, and entropy.

\subsection{Ensemble form}

Averaging \eqref{eq:Sigma_entropy_exact} over the forward ensemble gives
\begin{equation}
    \boxed{
    \avg{Y}-\Delta(\beta F)
    =\avg{\Delta s_{\tot}}
    +D_{\mathrm{KL}}\!
    \left(p_\tau\,\middle\|\,p_{\mathrm{eq}}^\tau\right),
    }
    \label{eq:ensemble_work_entropy_KL}
\end{equation}
where
\begin{equation}
    D_{\mathrm{KL}}\!
    \left(p_\tau\,\middle\|\,p_{\mathrm{eq}}^\tau\right)
    =\int\dd x\,p_\tau(x)
    \ln\frac{p_\tau(x)}{p_{\mathrm{eq}}^\tau(x)}\geq 0.
\end{equation}
Thus the mean dissipated reduced work consists of two nonnegative contributions: the mean total entropy production along the process and the relative entropy between the actual final distribution and the reference final canonical distribution.

If the actual final distribution equals the reference final canonical distribution, then the logarithmic mismatch term vanishes trajectory by trajectory, and \eqref{eq:Sigma_entropy_exact} reduces to
\begin{equation}
    \boxed{
    Y[x]=\Delta(\beta F)+\Delta s_{\tot}[x].
    }
    \label{eq:Y_betaF_entropy_simple}
\end{equation}
At constant temperature, \(Y=\beta W\) and \(\Delta(\beta F)=\beta\Delta F\), so the same relation becomes
\begin{equation}
    \boxed{
    \beta\bigl(W[x]-\Delta F\bigr)=\Delta s_{\tot}[x]
    }
    \label{eq:constant_beta_work_entropy}
\end{equation}
under the same final-distribution condition.  In the general case, the exact relation is \eqref{eq:Sigma_entropy_exact}, not \eqref{eq:constant_beta_work_entropy}.

\section{Application to score-based diffusion models}
\label{sec:diffusion_models}

This section explains how the preceding stochastic-thermodynamic structure can be used to model score-based diffusion generative models.  The goal is not to claim that a neural network literally performs mechanical work.  Rather, the claim is that a diffusion model defines a stochastic transport process in path space, and path-space irreversibility has the same mathematical structure as entropy production.

\subsection{Forward noising as a nonequilibrium stochastic process}

Consider a score-based forward noising process in \(d\) dimensions,
\begin{equation}
    \dd X_t=b_t(X_t)\dt+g_t\dd W_t,
    \qquad 0\leq t\leq T,
    \label{eq:dm_forward_sde}
\end{equation}
where \(X_0\sim p_{\mathrm{data}}\), \(W_t\) is a standard \(d\)-dimensional Wiener process, and the marginal density of \(X_t\) is denoted by \(p_t(x)\).  Usually the forward process is chosen so that
\begin{equation}
    p_T(x)\simeq \pi_T(x),
    \label{eq:dm_terminal_prior}
\end{equation}
where \(\pi_T\) is a simple prior, such as a standard Gaussian.

Equation \eqref{eq:dm_forward_sde} has diffusion matrix \(g_t^2 I\).  In the notation of the Langevin model used above,
\begin{equation}
    D_t=\frac{g_t^2}{2}.
    \label{eq:dm_D}
\end{equation}
If a mobility \(\mu_t\) is chosen, the same diffusion coefficient can be assigned an effective bath temperature
\begin{equation}
    T_t^{\mathrm{eff}}=\frac{D_t}{\mu_t}=\frac{g_t^2}{2\mu_t}.
    \label{eq:dm_effective_temperature}
\end{equation}
This identification is not unique, because only the product \(D_t=\mu_tT_t\) enters the stochastic dynamics.  Nevertheless, it gives a useful thermodynamic interpretation: the noise schedule \(g_t\) acts like a time-dependent thermal protocol.

\subsection{Exact reverse-time process}

Define the reverse-time variable
\begin{equation}
    \tau=T-t,
    \qquad
    Y_\tau=X_{T-\tau}.
    \label{eq:dm_reverse_time}
\end{equation}
The exact reverse-time diffusion has the same diffusion amplitude but a different drift:
\begin{equation}
    \dd Y_\tau=\widetilde b^*_\tau(Y_\tau)\dd\tau+g_{T-\tau}\dd \overline W_\tau,
    \label{eq:dm_exact_reverse_sde}
\end{equation}
with
\begin{equation}
    \boxed{
    \widetilde b^*_\tau(y)
    =-b_{T-\tau}(y)+g_{T-\tau}^2\nabla_y\ln p_{T-\tau}(y).
    }
    \label{eq:dm_exact_reverse_drift}
\end{equation}
The term
\begin{equation}
    \nabla_x\ln p_t(x)
    \label{eq:dm_true_score}
\end{equation}
 is the true score of the forward marginal distribution.  Therefore, exact generation from \(p_T\) back to \(p_0\) requires knowing the entire time-dependent score field.

\subsection{Learned reverse process and score error}

A diffusion model replaces the true score by a neural approximation
\begin{equation}
    s_\theta(x,t)\approx \nabla_x\ln p_t(x).
    \label{eq:dm_learned_score}
\end{equation}
The learned reverse drift is therefore
\begin{equation}
    \boxed{
    \widetilde b^\theta_\tau(y)
    =-b_{T-\tau}(y)+g_{T-\tau}^2s_\theta(y,T-\tau).
    }
    \label{eq:dm_learned_reverse_drift}
\end{equation}
Subtracting \eqref{eq:dm_learned_reverse_drift} from \eqref{eq:dm_exact_reverse_drift} gives
\begin{align}
    \widetilde b^*_\tau(y)-\widetilde b^\theta_\tau(y)
    &=g_{T-\tau}^2\nabla_y\ln p_{T-\tau}(y)
      -g_{T-\tau}^2s_\theta(y,T-\tau) \\
    &=g_{T-\tau}^2
    \left[\nabla_y\ln p_{T-\tau}(y)-s_\theta(y,T-\tau)\right].
    \label{eq:dm_drift_score_error}
\end{align}
Thus the score error is exactly the reverse-drift error divided by \(g_t^2\).

\subsection{Score error as path-space entropy production}
\label{subsec:score_error_entropy}

Let \(P_R^*\) denote the path measure of the exact reverse diffusion \eqref{eq:dm_exact_reverse_sde}, and let \(Q_R^\theta\) denote the path measure of the learned reverse diffusion \eqref{eq:dm_learned_reverse_drift}.  First assume that both reverse processes start from the same initial distribution, namely \(p_T\).  The two processes have the same diffusion coefficient and different drifts.  Girsanov's theorem gives the path-space relative entropy
\begin{equation}
    D_{\mathrm{KL}}(P_R^*\|Q_R^\theta)
    =\frac{1}{2}\int_0^T
    \mathbb E_{P_R^*}
    \left[
    \frac{\left\|\widetilde b^*_\tau(Y_\tau)-\widetilde b^\theta_\tau(Y_\tau)\right\|^2}
    {g_{T-\tau}^2}
    \right]\dd\tau .
    \label{eq:dm_girsanov_1}
\end{equation}
Substituting \eqref{eq:dm_drift_score_error},
\begin{align}
    &\frac{\left\|\widetilde b^*_\tau(y)-\widetilde b^\theta_\tau(y)\right\|^2}
    {g_{T-\tau}^2} \\
    &\qquad=
    \frac{\left\|g_{T-\tau}^2
    [\nabla_y\ln p_{T-\tau}(y)-s_\theta(y,T-\tau)]\right\|^2}
    {g_{T-\tau}^2} \\
    &\qquad=
    g_{T-\tau}^2
    \left\|s_\theta(y,T-\tau)-\nabla_y\ln p_{T-\tau}(y)\right\|^2 .
    \label{eq:dm_girsanov_substitution}
\end{align}
Under the exact reverse process, \(Y_\tau\sim p_{T-\tau}\).  Therefore
\begin{equation}
    D_{\mathrm{KL}}(P_R^*\|Q_R^\theta)
    =\frac{1}{2}\int_0^T
    g_{T-\tau}^2
    \mathbb E_{p_{T-\tau}}
    \left[
    \left\|s_\theta(x,T-\tau)-\nabla_x\ln p_{T-\tau}(x)\right\|^2
    \right]\dd\tau .
    \label{eq:dm_girsanov_2}
\end{equation}
Changing variables from reverse time \(\tau\) to forward time \(t=T-\tau\) gives
\begin{equation}
    \boxed{
    D_{\mathrm{KL}}(P_R^*\|Q_R^\theta)
    =\frac{1}{2}\int_0^T
    g_t^2
    \mathbb E_{p_t}
    \left[
    \left\|s_\theta(x,t)-\nabla_x\ln p_t(x)\right\|^2
    \right]\dt .
    }
    \label{eq:dm_score_KL}
\end{equation}
This is the precise sense in which score error becomes a path-space cost.

The connection to entropy production follows from the stochastic-thermodynamic identity that entropy production is a log-ratio of forward and reversed path probabilities.  Define a model-dependent path entropy production by
\begin{equation}
    \boxed{
    \Sigma_\theta[X]
    =\ln\frac{\dd P_F[X]}{\dd Q_R^{\theta,\Theta}[X]},
    }
    \label{eq:dm_model_entropy_def}
\end{equation}
where \(P_F\) is the forward noising path measure and \(Q_R^{\theta,\Theta}\) is the learned reverse path measure after time reversal.  If \(P_R^*\) is the exact time reversal of \(P_F\), then
\begin{equation}
    P_F=P_R^{*,\Theta}.
    \label{eq:dm_exact_time_reversal_measure}
\end{equation}
Averaging \eqref{eq:dm_model_entropy_def} over \(P_F\) yields
\begin{align}
    \mathbb E_{P_F}[\Sigma_\theta]
    &=D_{\mathrm{KL}}(P_F\|Q_R^{\theta,\Theta}) \\
    &=D_{\mathrm{KL}}(P_R^*\|Q_R^\theta).
    \label{eq:dm_entropy_KL_equivalence}
\end{align}
Combining \eqref{eq:dm_entropy_KL_equivalence} with \eqref{eq:dm_score_KL},
\begin{equation}
    \boxed{
    \mathbb E_{P_F}[\Sigma_\theta]
    =\frac{1}{2}\int_0^T
    g_t^2
    \mathbb E_{p_t}
    \left[
    \left\|s_\theta(x,t)-\nabla_x\ln p_t(x)\right\|^2
    \right]\dt .
    }
    \label{eq:dm_score_entropy_final}
\end{equation}
Thus, minimizing the weighted score-matching error is equivalent to minimizing a model-induced entropy production in path space, up to the precise choice of training weights.

\subsection{Endpoint mismatch}

The previous derivation assumed that the learned reverse process begins from the true terminal distribution \(p_T\).  In actual generation, the model often starts from a chosen prior \(\pi_T\).  If \(p_T\neq \pi_T\), the chain rule for path-space relative entropy gives an additional endpoint mismatch term:
\begin{equation}
    \boxed{
    \mathbb E_{P_F}[\Sigma_\theta]
    =D_{\mathrm{KL}}(p_T\|\pi_T)
    +\frac{1}{2}\int_0^T
    g_t^2
    \mathbb E_{p_t}
    \left[
    \left\|s_\theta(x,t)-\nabla_x\ln p_t(x)\right\|^2
    \right]\dt .
    }
    \label{eq:dm_endpoint_plus_score}
\end{equation}
Therefore the thermodynamic irreversibility of a diffusion model has two sources:
\begin{equation}
    \boxed{
    \text{terminal prior mismatch}
    +
    \text{score-induced path mismatch}.
    }
    \label{eq:dm_two_sources}
\end{equation}

\subsection{Work as a model-quality measure}

The preceding section shows that model error can be measured by path-space entropy production.  If a reduced potential
\begin{equation}
    u_t(x)=\beta_tV_t(x)
    \label{eq:dm_reduced_potential}
\end{equation}
 is specified, then the work-free-energy relation derived above gives a second, work-based diagnostic.  Along a generation path \(\Gamma=\{X_t\}_{0\leq t\leq T}\), define the reduced work functional
\begin{equation}
    Y_\theta[\Gamma]
    =\int_0^T \partial_tu_t(X_t)\dt
    +\int_0^T \beta_t f_\theta(X_t,t)\Strat\dd X_t,
    \label{eq:dm_Ytheta}
\end{equation}
where \(f_\theta\) denotes any nonconservative control component induced by the learned reverse dynamics.  The reference reduced free-energy difference is
\begin{equation}
    \Delta\mathcal F
    =-\ln Z_T+\ln Z_0,
    \qquad
    Z_t=\int \dd x\,e^{-u_t(x)}.
    \label{eq:dm_delta_mathcalF}
\end{equation}
Then a work-based inefficiency can be defined as
\begin{equation}
    \boxed{
    \mathcal L_{\mathrm{work}}(\theta)
    =\mathbb E_{Q_\theta}[Y_\theta]-\Delta\mathcal F.
    }
    \label{eq:dm_work_quality}
\end{equation}
In the ideal reversible limit this quantity is minimized.  In practical diffusion models it is increased by score error, finite-time driving, discretization error, and mismatch between the terminal noising distribution and the chosen prior.  This suggests a thermodynamic criterion for model quality:
\begin{equation}
    \boxed{
    \text{a better generative model realizes the prior-to-data transport with smaller excess work.}
    }
    \label{eq:dm_better_model_excess_work}
\end{equation}

The distribution of work values can also be informative.  A broad or heavy-tailed distribution of \(Y_\theta\) indicates that some generated trajectories require unusually large control effort.  A time-resolved version of \eqref{eq:dm_work_quality} can diagnose which noise levels contribute most strongly to dissipation and can therefore guide time-step allocation, noise-schedule design, or model-capacity allocation.

\subsection{Diffusion model as a thermodynamic machine}

A standard diffusion model is not a heat engine in the strict thermodynamic sense, because it is not naturally a cyclic device that extracts mechanical work from heat reservoirs.  It is better viewed as a work-consuming information machine.  The reverse generation process transforms a high-entropy prior into structured data:
\begin{equation}
    \pi_T \longrightarrow p_{\mathrm{data}}.
    \label{eq:dm_prior_to_data}
\end{equation}
This resembles an information refrigerator or heat pump: entropy is reduced in the sample distribution, and the required control is supplied by the learned score field.

One can nevertheless build a cycle by composing forward noising and learned reverse generation:
\begin{equation}
    p_{\mathrm{data}}\xrightarrow{\text{forward noising}}p_T
    \xrightarrow{\text{learned reverse}}p_\theta.
    \label{eq:dm_cycle}
\end{equation}
The cycle irreversibility is measured by
\begin{equation}
    \boxed{
    \mathcal H_\theta
    =D_{\mathrm{KL}}(P_F\|Q_R^{\theta,\Theta})
    =\mathbb E_{P_F}[\Sigma_\theta].
    }
    \label{eq:dm_cycle_hysteresis}
\end{equation}
This quantity can be interpreted as a thermodynamic hysteresis of the generative cycle.  If the learned reverse process exactly inverts the forward process, then \(\mathcal H_\theta=0\).  If the reverse process is imperfect, then \(\mathcal H_\theta>0\).

\begin{tcolorbox}[colback=green!3!white,colframe=green!40!black,title=Main interpretation for diffusion models]
A diffusion model can be understood as a nonequilibrium stochastic transport machine.  The score network supplies the control field that attempts to reverse the forward noising process.  Score error produces a reverse-drift mismatch, which produces a path-space KL divergence, which is a model-induced entropy production.  When a reduced potential is specified, the same structure can be written as excess work above a reduced free-energy difference.
\end{tcolorbox}

\section{Ito representation of the thermodynamic line integrals}

The SDE \eqref{eq:langevin_sde} has additive noise because \(D_t\) is independent of \(x\).  Thus the Ito and Stratonovich forms of the Langevin equation have the same drift.  However, thermodynamic heat and nonconservative work are line integrals and are defined above in the Stratonovich sense.

For any smooth function \(A(x,t)\), the conversion rule is
\begin{equation}
    \int_0^\tau A(x_t,t)\Strat\dd x_t
    =\int_0^\tau A(x_t,t)\dd x_t
    +\int_0^\tau D_t\partial_xA(x_t,t)\dt,
    \label{eq:ito_conversion}
\end{equation}
where the integral on the right-hand side is Ito.  Therefore
\begin{equation}
    Q[x]=\int_0^\tau F\Strat\dd x_t
    =\int_0^\tau F\dd x_t+
    \int_0^\tau D_t\partial_xF\dt.
    \label{eq:heat_ito}
\end{equation}
Similarly,
\begin{equation}
    \int_0^\tau f\Strat\dd x_t
    =\int_0^\tau f\dd x_t+
    \int_0^\tau D_t\partial_xf\dt.
    \label{eq:fwork_ito}
\end{equation}
The terms \(\int \partial_\lambda V\dot\lambda\dt\) and \(\int\dot\beta V\dt\) are ordinary time integrals and have no Ito-Stratonovich ambiguity.

\section{Summary of definitions and identities}

For
\begin{equation}
    \dd x_t=\mu_tF(x_t,\lambda_t)\dt+\sqrt{2D_t}\,\dd W_t,
    \qquad
    D_t=\mu_tT_t=\frac{\mu_t}{\beta_t},
\end{equation}
with
\begin{equation}
    F(x,\lambda)=-\partial_xV(x,\lambda)+f(x,\lambda),
\end{equation}
the trajectory-level quantities are:

\begin{tcolorbox}[colback=blue!3!white,colframe=blue!50!black,title=Definitions]
\begin{align}
    E_t&=V(x_t,\lambda_t), \\
    \delta w_t&=\partial_\lambda V(x_t,\lambda_t)\dot\lambda_t\dt
    +f(x_t,\lambda_t)\Strat\dd x_t, \\
    \delta q_t&=F(x_t,\lambda_t)\Strat\dd x_t, \\
    s_{\sys}(t)&=-\ln p(x_t,t), \\
    \Delta s_{\med}[x]&=\int_0^\tau\beta_tF(x_t,\lambda_t)\Strat\dd x_t, \\
    \Delta s_{\tot}[x]&=\Delta s_{\med}[x]
    -\ln p(x_\tau,\tau)+\ln p_0(x_0).
\end{align}
\end{tcolorbox}

They satisfy the trajectory first law
\begin{equation}
    \boxed{\delta w_t=\dd E_t+\delta q_t}
\end{equation}
and the integral fluctuation theorem
\begin{equation}
    \boxed{\avg{e^{-\Delta s_{\tot}}}=1.}
\end{equation}

For canonical initial and reference final distributions generated by \(V\), define
\begin{equation}
    Z(\beta,\lambda)=\int\dd x\,e^{-\beta V(x,\lambda)},
    \qquad
    F(\beta,\lambda)=-\beta^{-1}\ln Z(\beta,\lambda).
\end{equation}
Then the reduced work functional is
\begin{equation}
    \boxed{
    Y[x]
    =\int_0^\tau\beta_t\partial_\lambda V(x_t,\lambda_t)\dot\lambda_t\dt
    +\int_0^\tau\beta_tf(x_t,\lambda_t)\Strat\dd x_t
    +\int_0^\tau\dot\beta_tV(x_t,\lambda_t)\dt .
    }
\end{equation}
The generalized Jarzynski identity is
\begin{equation}
    \boxed{
    \avg{e^{-Y[x]}}
    =\frac{Z(\beta_\tau,\lambda_\tau)}{Z(\beta_0,\lambda_0)}
    =\exp\left[-\beta_\tau F(\beta_\tau,\lambda_\tau)
    +\beta_0F(\beta_0,\lambda_0)\right].
    }
\end{equation}

The dissipated reduced work is
\begin{equation}
    \boxed{
    \Sigma[x]=Y[x]-\Delta(\beta F),
    \qquad
    \avg{e^{-\Sigma[x]}}=1.
    }
\end{equation}
For the actual final distribution \(p_\tau(x)\) and the reference final canonical distribution
\begin{equation}
    p_{\mathrm{eq}}^\tau(x)=\frac{e^{-\beta_\tau V(x,\lambda_\tau)}}{Z_\tau},
\end{equation}
the exact relation to entropy production is
\begin{equation}
    \boxed{
    Y[x]-\Delta(\beta F)
    =\Delta s_{\tot}[x]
    +\ln\frac{p_\tau(x_\tau)}{p_{\mathrm{eq}}^\tau(x_\tau)}.
    }
\end{equation}
Consequently,
\begin{equation}
    \boxed{
    \avg{Y}-\Delta(\beta F)
    =\avg{\Delta s_{\tot}}
    +D_{\mathrm{KL}}\!\left(p_\tau\,\middle\|\,p_{\mathrm{eq}}^\tau\right).
    }
\end{equation}

\end{document}